# Broken weak and strong spin rotational symmetries and tunable interaction between phonon and the continuum in $Cr_2Ge_2Te_6$

Atul G. Chakkar[*], Deepu Kumar, Pradeep Kumar[#]

*School of Physical Sciences, Indian Institute of Technology Mandi, 175005, India*

**Abstract**

Phase transitions with lowering temperature is a manifestation of decreased entropy and within the Landau theoretical framework these are accompanied by the symmetry breaking. Whenever a symmetry is broken weakly or strongly, it leaves its trail and the same may be captured indirectly using renormalization of the quasi-particle excitations. $Cr_2Ge_2Te_6$, a quasi-two-dimensional magnetic material, provides a rich playground to probe dynamics of the quasi-particle excitations as well as multiple phase transitions with lowering temperature intimately linked with the lattice and spin degrees of freedom. Here, we report in-depth inelastic light scattering measurements on single crystals of $Cr_2Ge_2Te_6$ as a function of temperature, from 6 K to 330 K, and polarization. Our measurements reveal the long as well as short range ordering of the spins below $T_c$ (~ 60 K) and $T^*$ (~ 180 K), respectively; setting the stage for broken rotational and time reversal symmetry, gauged via the distinct renormalization of the phonon self-energy parameters along with the modes intensity. Our measurements also uncovered an intriguing dependence of the interaction strength between discrete state (phonon here) and the underlying continuum, quantified using the Fano asymmetry parameter, as a function of the scattered light polarization. Our results suggest the possibility of tuning the interaction strength using controlled scattered light and symmetry in this 2D magnet.

[*]E-mail: atulchakkar16@gmail.com
[#]E-mail: pkumar@iitmandi.ac.in

# 1. Introduction

Two-dimensional (2D) Transition Metal Tri-Chalcogenides (TMTCs) belong to the class of van der Waals layered materials. Among these, a family of $Cr_2X_2Te_6$ (X= Si, Ge) are intrinsic ferromagnetic semiconductors. $Cr_2Ge_2Te_6$ is a 2D Heisenberg layered van der Waals ferromagnetic semiconductor (0.74 eV; indirect band gap) with curie temperature in the range of ~ 55-65 K [1–6]. On the other hand, a sister compound $Cr_2Si_2Te_6$ is an Ising ferromagnetic semiconductor (0.4 eV; direct band gap) with a curie temperature of 32 K [4,7,8]. As these materials exhibit ferromagnetism and are semiconducting in nature, these are potential candidates for the next-generation spintronic devices, nano-electronics, and are useful as a substrate in ferromagnetic insulator-topological insulator heterostructures. In the hexagonal unit cell, three layers of $Cr_2Ge_2Te_6$ are stacked in an *ABC* sequence along the *c*-axis. The Cr atom is at the center and surrounded by six Te atoms in a distorted octahedron [4]. The origin of ferromagnetism in $Cr_2Ge_2Te_6$ is attributed to the super-exchange mechanism between the Cr-Te-Cr magnetic ion having spin S=3/2 (Cr - $3d^3$) with high spin configuration [6,9,10]. Spin-phonon and electron-phonon coupling both play an important role in many properties such as spin dynamics and transport properties. According to the Mermin-Wagner theorem, in a 2D Van der Waals system, long-range magnetic order is affected by thermal fluctuations and these fluctuations may be counteracted by magnetic anisotropy [2,11]. For 2D ferromagnetic $Cr_2Ge_2Te_6$, a Heisenberg Hamiltonian with magnetic anisotropies can be given as: $H = \sum_{i,j} J_{ij} S_i \cdot S_j + \sum_i A(S_i^j)^2$, where '$S_i$' is the spin operator on site i, $J_{ij}$ is the exchange interaction between sites i and j sites, A is the single ion anisotropy. This magnetic anisotropy establishes ferromagnetic order in 2D materials at finite temperatures with the continuous rotational symmetry breaking of the Hamiltonian [2]. In $Cr_2Ge_2Te_6$, to understand the nature of phase transitions i.e. long and short-range magnetic ordering, and to explore the dynamics of quasi-particle excitations and their interactions, temperature-dependent Raman spectroscopic investigation may provide deeper understanding dynamics of such interactions like anharmonicity originating from the phonon-phonon interaction, electron-phonon coupling and quasi-particle dynamics. For the current system under probe, a long-range ordering of spins has been reported in the temperature range of ~ 55-65 K [6,12]. Additionally, a short-range ordering has also been suggested around ~ 160-200 K [13–15]. Therefore, it is important to investigate and understand such quasi-particle excitations and magnetic ordering using in-depth temperature dependent Raman studies. Raman spectroscopy is a powerful technique to study the different phenomena in 2D materials such as quantum spin liquids, transition metal chalcogenides, and other magnetic materials [16–21].

In this work, we have performed comprehensive Raman spectroscopic measurements in the temperature range of ~ 6 to 330 K covering a broad spectral range from 5 to 2500 cm$^{-1}$ to understand the important phenomena in this 2D magnetic $Cr_2Ge_2Te_6$. For a single crystal of $Cr_2Ge_2Te_6$, we observed a Raman signature of ferromagnetic transition at $T_C \sim 60 K$ and the short-range spin-spin correlation is also observed around $T^* \sim 180 K$ reflected in the renormalized self-energy parameters of the Raman active phonon modes. We also studied the interaction between quasi-particle excitations such as phonons, spins, and electronic and/or magnetic continuum as a function of temperature and this also revealed the magnetic transition effects at the above-mentioned transition temperatures (i.e. $T_C$ and $T^*$). Our studies also uncovered the potential route to control the interaction strength between phonons and the underlying continuum in this 2D magnetic material.

## 2. Experimental details

Temperature dependent Raman scattering measurements were performed using the LabRAM HR-Evolution Raman spectrometer in the backscattering configuration and closed-cycle He-flow cryostat (Montana) from 6 to 330 K with ± 0.1K accuracy. The spectra were excited using a 532 nm laser and the laser power was kept very low (< 0.2 mW) to avoid the local heating on the sample. A 50x long working distance objective was used to focus the laser light on the sample as well as to collect the scattered light from the sample. Polarization-dependent Raman measurements were also performed at five different temperatures below and above long-range ($T_C$) magnetic ordering i.e. 6, 45 150, 250, and 300 K.

## 3. Results and Discussions

### 3.1. Lattice Vibrations in $Cr_2Ge_2Te_6$

Bulk single crystal of $Cr_2Ge_2Te_6$ belongs to the point group $C_{3i}$ (space group $R\bar{3}$, #148) and have hexagonal structure [1,3,6]. Bulk $Cr_2Ge_2Te_6$ unit cell consists of 10 atoms per unit cell which gives rise to 30 phonon branches at $\Gamma$ point of the Brillouin zone, and can be expressed by the following irreducible representation as $\Gamma = 5A_g + 5A_u + 5E_g + 5E_u$. There are 27 optical and 3 acoustical branches with the irreducible representation $\Gamma_{optical} = 5A_g + 4A_u + 5E_g + 4E_u$ and $A_u + E_u$, respectively. From the 27 optical branches, 15 are Raman active; $\Gamma_{Raman} = 5A_g + 5E_g$, whereas remaining 12 are infrared active; $\Gamma_{IR} = 4A_u + 4E_u$. The vibrations of phonon modes corresponds to the symmetry $E_g$ and $A_g$ are reported by Y. Sun et al [22]. Figure 1(a) shows the Raman spectrum of $Cr_2Ge_2Te_6$ in the spectral range of 65-500 cm$^{-1}$ recorded at 6 K. Spectra are fitted using a sum of Lorentzian functions to extract the self-energy parameters for the phonon modes i.e. mode frequency (ω), and full width at half maximum (FWHM), as well as the intensity. We observed eight Raman active phonon modes at 6 K, for convenience named as P1(~ 95 cm$^{-1}$), P2( ~ 106 cm$^-$

$^{1}$), P3(~ 126 cm$^{-1}$), P4(~ 144 cm$^{-1}$), P5(~ 231 cm$^{-1}$), P6(~ 276 cm$^{-1}$), P7(~ 292 cm$^{-1}$) and P8(~ 446 cm$^{-1}$). We also recorded the Raman spectra in the higher frequency range up to 2500 cm$^{-1}$, but did not observed any signature of the Raman active excitation (see inset in Fig.1(b)). The spectrum consists of very strong characteristic phonon modes P3 and P4 located at ~ 126 and 144 cm$^{-1}$. Mode P3 shows the asymmetric line shape nature and it is discussed in detail in section 3.3. Phonon modes P1, P6 and P2, P3 are assigned $E_{1g}$ and $E_{2g}$ symmetry, respectively; while modes P4, P8 are assigned $A_g$ symmetry [6,23]. Figure 1(b) shows temperature evolution of the Raman spectrum of Cr$_2$Ge$_2$Te$_6$ in the range of 6 to 330 K. With increasing the temperature from 6 K to 330 K, peak frequencies are red-shifted and peaks get broadened. Peak P2 becomes very weak with increasing the temperature. The quasi-elastic scattering in the low-frequency region of the spectra also evolves with the temperature.

### 3.2. Temperature dependence of the phonon modes

### 3.2.1. Phonon anharmonicity, thermal expansion, and spin-phonon coupling

Figure 2(a)-(d), shows the temperature-dependent frequency and linewidth of the observed prominent phonon modes i.e. P1-P4, P6, and P8. The following observation can be made: (i) All the observed phonon modes show hardening with lowering the temperature from 330 K to ~ 60 K. Below ~ 60 K, the frequency of modes P1, P3, P4, P6, and P8 show an upward trend upon entering the long-range magnetic ordering phase. We attribute this to the effect of spin-phonon coupling below ~ 60 K. (ii) FWHM of the mode P1 decreases with lowering the temperature. Interestingly, phonon modes P3, P4, P6, and P8 show a decrease in FWHM with a decreasing the temperature up to ~ 60 K. However, below this temperature, FWHM remains nearly constant for the modes P3 and P8 and increases for P4 and P6 with further lowering the temperature. The increases in linewidth below ~ 60 K may be understood invoking emergence of the magnetic decay channel.

In the long-range ordered phase, magnetic channel becomes active and as a result optical phonons will have more channels to decay and leads to shorter lifetime ($\tau$). This may ultimately lead to increase in the linwidth ($\propto \sim 1/\tau$) below $T_C$ as seen in our observations. FWHM of the weak mode P2 is observed to be nearly constant in the full temperature range. We observed that frequency decreases with increasing temperature (from ~ 60 -330 K) for the modes P1, P2, P3, P4, and P6, the change is ~ 3.5 %, 4 %, 2.9 %, 2 %, 2.4 %, respectively, with respect to the frequency at 60 K. For the mode P8, the frequency change is ~ 0.46 % with respect to the frequency at 60 K. The effect of short-range ordering of the spins, which start much above the long-range ordering temperature, may also be captured via renormalization of the self-energy parameters of the phonons as well as the intensity. For $Cr_2Ge_2Te_6$, it has been reported in the transport studies that short range ordering starts building up as high as ~ 200K. We observed a clear change in the mode frequencies and FWHM in the vicinity of 180K, in particular a change in the slope or a kink is observed around this temperature clearly visible in the modes P3, P4, and P8 (see Fig. 2). We attributed this to the effects of short-range ordering of the spins. Building up of the short-range ordering may affects the Raman cross-section and the same may be captured via the phonon modes intensity evolution. The effect of short-range ordering is also seen in the temperature evolution of the phonon modes intensity. It is advocated that in case of magnetic materials, the long-range as well as the short-range ordering may affect the phonon mode intensities [24,25]. Figure 2(e) shows the temperature dependence of the phonon mode intensities for the modes P1-P4, and P8. Intensity of the mode P1 increases slowly with increasing temperature up to ~ 180 K and above ~ 180 K, the slope is changed and intensity increases sharply. For the phonon mode P2, intensity decreases with increasing temperature in the full temperature range, though there is a change in slope around 60 and 180 K. Intensity of the mode P3 and P4 decreases with increasing temperature up to ~ 180

K, above ~ 180 K there is a change in slope and intensity increases with further increase in the temperature. Intensity of the mode P8 decreases slightly till ~ 60 K, remains nearly constant with further increase in the temperature up to ~ 180 K, and after that it increases sharply till 330 K. These changes in intensity of the phonon modes around ~ 60 K and 180 K are clearly the manifestation of the long and short-range spin ordering.

Generally, temperature evolution of the phonon modes may have contribution from three factors: (1) anharmonic effect, $\Delta\omega_{anh}(T)$, (2) thermal expansion of the lattice, $\Delta\omega_{latt}(T)$ and (3) spin-phonon coupling, $\Delta\omega_{sp-ph}(T)$. Change in the phonon mode frequencies from the above three factors may be given as [26,27]:

$$\Delta\omega(T) = \Delta\omega_{anh}(T) + \Delta\omega_{latt}(T) + \Delta\omega_{sp-ph}(T) \tag{1}$$

Anharmonic temperature dependence of the frequency and FWHM of the phonon modes may be understood by considering the three-phonon anharmonic process, decaying of an optical phonon into two equivalent acoustics phonons with equal frequency but opposite momentum, suggesting by Klemens and may be written as [28]:

$$\Delta\omega_{anh} = \omega(T) - \omega_0 = A\left(1 + \frac{2}{e^x - 1}\right) \tag{2}$$

$$\Delta\Gamma_{anh} = \Gamma(T) - \Gamma_0 = C\left(1 + \frac{2}{e^x - 1}\right) \tag{3}$$

where $\omega_0$ and $\Gamma_0$ are the frequency and FWHM at 0 K, respectively; and $x = \frac{\hbar\omega_0}{2k_B T}$. $A$ and $C$ are the constant parameters associated with the change in frequency and FWHM with temperature, respectively. The solid red lines for the case of frequency and FWHM in a temperature range of 60 to 330 K in Figure 2(a)-(d) are fitted curves using the above equations (2) and (3), respectively.

The fitting is in quite good agreement with the experimental data. The best fit parameters are listed in Table-I.

The second term in equation (1) corresponds to the thermal expansion of the lattice, also known as the quasi-harmonic effect and corresponding change in the frequency may be given as:

$$\Delta\omega_{latt}(T) = \omega_0 \left\{ \exp\left[-3\gamma \int_{T_0}^{T} \alpha(T) dT - 1\right] \right\} \quad (4)$$

where, $\gamma$ is the Gruneisen parameter and $\alpha(T)$ is the linear thermal expansion coefficient (TEC). For convenience, we have written the product of $\gamma$ and $\alpha(T)$ as a function of temperature:

$$\gamma\alpha(T) = b_0 + b_1 T + b_2 T + ..... \quad (5)$$

where $b_0, b_1, b_2$ are constant (for the values see supplementary Table-S1). For the case of Cr$_2$Ge$_2$Te$_6$, the Gruneisen parameter ($\gamma$) for phonon modes is not known. Generally, $\gamma$ is of the order of 1-3, hence we have taken three different values i.e., $\gamma = 1, 2, 3$ for each phonon mode. Figure 3(a) and (b) shows the temperature dependence of the extracted linear thermal expansion coefficient, $\alpha(T)$, for the phonon modes P1-P4, P6, and P8 in the temperature range of ~ 60-330 K. Modes P1 and P4 show an increase in the TEC with an increase in temperature. The phonon mode P2 shows the slight decreases in TEC with increase in the temperature up to temperature ~ 100 K and above 100 K, it increases sharply. The phonon modes P3, and P6 shows the decrease in the TEC with increasing temperature up to ~ 210 K and increase with further increase in the temperature. Mode P8 shows a decrease in TEC with increasing the temperature up to ~ 150 K, and it increases with further increasing the temperature till 330 K.

To understand the effect of anharmonicity in the full temperature range of ~ 6-330 K, we extrapolated the anharmonic model up to the lowest recorded temperature i.e. 6 K. Interestingly one could clearly see a deviation of this anharmonic phonon-phonon fit from the experimental data, see the solid green lines in Figure 2(a)-(d). This suggests that alone anharmonic model cannot capture the entire picture to understand the temperature dependence of the phonon modes. Further, one could see that the deviation is starting from ~ 60 K, which is close to the reported $T_C$ for the case of Cr$_2$Ge$_2$Te$_6$. The observed anomalous temperature dependence of the phonon modes below 60 K may be understood by considering the coupling of phonons with magnetic degree of freedom. Below $T_C$, the spin-phonon coupling (SPC) in magnetic materials is observed to play a significant role in renormalization of the frequency and linewidth of the phonon modes. Temperature-dependent frequency could be the best parameter to gauge role of the SPC [29–31].

Generally, below $T_C$ spins get ordered and the spin-phonon interaction comes into the picture. Spin-phonon coupling describe the interaction between the lattice degree of freedom (phonons) and the underlying magnetic degree of freedom (spins). Keeping only the dominant contribution, the Hamiltonian for the spin-phonon interaction may be given as $H = \sum_{i,j} J_{ij} \left( \vec{S_i} \cdot \vec{S_j} \right)$, where $J_{ij}$ is the exchange coupling integral [32]. The phonon frequency is sensitive to correlations of the spins and the change in frequencies due to the SPC may be given as [21,31,32];

$\Delta \omega_{sp-ph} = \omega_{ph} - \omega_{ph}^0 = \lambda * \left\langle \vec{S_i} \cdot \vec{S_j} \right\rangle$, where $\omega_{ph}$ is the phonon frequency and $\omega_{ph}^0$ is the bare phonon frequency. $\lambda$ is the spin-phonon coupling coefficient which describe coupling strength between spins and phonons and its value can be both positive (negative) for the hardening (softening) of the phonons below $T_C$. Further $\lambda$ may be derived from the exchange coupling integral $J_{ij}$ as

$\lambda = \frac{\partial^2 J_{ij}}{\partial r^2}$, where r is the atomic displacement. $\left\langle \vec{S_i} \cdot \vec{S_j} \right\rangle$ is the spin-spin correlation function between the i$^{th}$ and j$^{th}$ site of the magnetic ions and within the mean field approximation it may be given as $\left\langle \vec{S_i} \cdot \vec{S_j} \right\rangle = -S^2 \phi(T)$, where $S$ is spin on magnetic ion/site and here we have S=3/2 for the Cr magnetic ion i.e. $Cr^{3+}$ [6,9,10]. Therefore, the temperature-dependent frequency of the phonon considering the contributions from the SPC can be written as;

$$\omega_{sp-ph}(T) = \omega_{ph}^0 - \lambda * S^2 * \phi(T) \qquad (6)$$

and $\phi(T)$ can be written as $\phi(T) = 1 - \left(\dfrac{T}{T_C}\right)^\eta$, where $\eta$ is the critical exponent. $\phi(T)$ is the order parameter. It becomes zero at the curie temperature ($T_C$) and unity at absolute zero temperature while above the curie temperature $T > T_C$ it is zero [18,31,32]. Insets in Fig. 2, show the temperature dependence of the modes frequency below $T_C$ i.e. ~ 60 to 6 K, where solid blue lines are the fitted curves using equation 6. Here, it should be noted that we have fixed the $T_C$ (60 K). The best fit parameters are listed in Table-II. A large value of the extracted $\lambda$ reflects strong coupling between phonons and spins while negative value indicates the hardening of frequency of the phonon modes with decreasing the temperature up to 6 K.

### 3.3. Fano asymmetry and its polarization dependence

In the Figure 1(b), phonon mode P3 shows the right-handed asymmetric nature, and the origin of this asymmetry may be captured by the Fano model [33–35]. The Raman asymmetric line-shape broadening occurs due to the interaction between the electronic or magnetic continuum with the discrete quasi-particle excitations (phonons here) and this interaction may result into the asymmetric line shape, termed as Fano asymmetry. Underlying origin of this effect is the spin dependent electron polarizability involving both spin-photon and spin-phonon coupling [33,36]. This interaction plays a crucial role in the semiconducting materials because it affects the electrical and optical properties of the semiconductors and such study may be useful in different applications [37]. The Raman line-shape broadening due to the Fano interaction is given as

$$I(\omega) \propto \frac{(1+\delta/q)^2}{1+\delta^2}, \text{ where } \delta = \frac{\omega-\omega_0}{\Gamma}$$, and $I(\omega)$ is the Raman intensity as a function of frequency $\omega$. $\omega_0$ and $\Gamma$ are the frequency and FWHM of the bare /uncoupled phonon, respectively. $q$ is the asymmetry parameter that characterizes the coupling strength between the phonon and the continuum, quantified by the parameter $1/|q|$. Microscopically $q$ may be defined as:

$$|q|^2 = \frac{|\langle\phi|T|i\rangle|^2}{|\langle\psi_E|T|i\rangle|^2 .\pi^2 |V_E|^2}$$ ; it represents ratio of the transition probability to the modified discrete state, $|\phi\rangle$, and to a band width, $2\pi|V_E|^2$, of the unperturbed continuum states, $|\psi_E\rangle$. T is some transition operator between initial state $|i\rangle$ and the state $|\phi\rangle$ and $|\psi_E\rangle$. $V_E(=\langle\psi_E|H|\phi\rangle)$ is some transition tells the interaction between the discrete state and the continuum. Therefore, in the limit $1/|q|(\propto |V_E|) \to \infty$; coupling is stronger and causes the peak to be more asymmetric while in the limit $1/|q| \to 0$; coupling is weak and this results into the Lorentzian line shape [21,34,37–40]. Figure 4(b) shows the temperature dependence of the coupling strength ($1/|q|$) for the Fano peak (P3) in the temperature range of 6-330 K. We note that with lowering the temperature $1/|q|$ increases continuously and below $T_C (\sim 60K)$, there is a sharp rise in its magnitude reflecting the strong coupling between phonons and the underlying magnetic degree of freedom.

Asymmetry parameter, $q$, is directly linked with the coupling strength and is intimately connected with the electronic polarizability. Since electronic polarizability have its dependencies on the spins as: $\chi_{ab} = \chi_{ab}^0 + \sum \chi_{ab}(\sigma;\tau,m)S_\sigma(\tau,m) + \sum \chi_{ab}(\sigma;\tau,m;\sigma';\tau',m')S_\sigma(\tau,m)S_{\sigma'}(\tau',m') + ....$ ; Therefore, its detailed study may reveal the coupling strength and its potential variation as a function of the light polarization may uncover its intrinsic dependence on the spins and photons as also advocated in some of the earliest studies on Fano asymmetry [33,35,36]. This may pave the way for controlling the interaction strength as a function of the light polarization. This may also open the possibility of potential quantum sensor such as quantifying interaction strength as a function of the photon polarization.

To probe the possible dependence of the asymmetry parameter, $q$, as a function of the scattered light polarization, we performed detailed polarization dependence measurements both below and above the long-range ordering temperature. As $1/|q|$ reflects the interaction strength, the maxima in $1/|q|$ at a particular scattered light polarization may be picked up and the variation in the maxima as a function of temperature may also be observed underlying the effects of the spins and photons. Figure 4(c) show the asymmetry parameter $1/|q|$ as a function of polarization angle ($\theta$) at five different temperatures 6 K, 45 K below $T_C$ and 150 K, 250 K, 300 K above $T_C$. We note that below $T_C$, $1/|q|$ show maxima at the scattering angles of ~ 80 and ~ 260 degrees at 6 and 45 K, respectively. However, above $T_C$ (i.e. 150, 250, and 300 K) it shows the maxima at ~100 and ~280 degrees. The maxima at different scattering angles above and below $T_C$ reflects the effect of long-range magnetic ordering on the coupling between the phonon and the underlying continuum. It also shows the coupling strength dependence on the scattering angle of the electric field vector. From Fig. 4(c), the coupling strength ($1/|q|$) shows the minimum value for the parallel polarized configuration (XX), while it shows the maximum value for the cross-polarized configuration (XY). This observation suggests that coupling strength may be controlled using polarization of the light as advocated. More similar studies on these quasi 2D magnetic systems may be helpful to fully understand the underlying mechanism for this and realizing its full potential. A quantitative understanding of this maxima in asymmetry parameter calls for an in-depth theoretical study. Our observations also highlight the prospect to control the quantum pathways of the inelastically scattered light and this may provide a way in future to control these pathways for use in quantum technology.

**3.4 Quasi-elastic scattering**

In magnetic materials, the spin energy fluctuation or fluctuations of the magnetic energy density causes quasi-elastic Raman scattering and this shows a notion of the magnetic contribution to the specific heat. The magnetic specific heat ($C_m$) provides an idea about underlying magnetic excitation, spin fluctuations, and Raman signature of the change of phase with a temperature change [41-43]. Raman response, $\chi''(\omega)$, is an imaginary part of the susceptibility obtained from the raw Raman intensity and is given as $I(\omega) \propto \chi''(\omega)[n(\omega)+1]$, where $n(\omega)$ is the Bose thermal factor. Figure 5(a) shows the Raman response at different temperatures in the spectral range of ~ 4 to 75 cm$^{-1}$. At high temperatures the Raman response is nearly flat and start building up only at low temperatures, reflecting more fluctuations in the magnetic excitations as compared to the high-temperature range. The Raman conductivity is another important parameter and is extracted from the Raman response by simply dividing by a Raman shift and given as: $\chi''(\omega)/\omega$. Figure 5(b) shows the Raman conductivity for different temperatures, and this is used to extract the dynamic Raman susceptibility $\chi^{dyn}$ and is given by the Kramers-Kronig relation:

$$\chi^{dyn} = \lim_{\omega \to 0} \chi(k=0,\omega) \equiv \frac{2}{\pi} \int_0^\infty \frac{\chi''(\omega)}{\omega} d\omega \qquad (7)$$

The upper cutoff value of frequency is taken as 75 cm$^{-1}$. Figure 5(c) shows the normalized dynamic Raman susceptibility $\chi^{dyn}$ as a function of temperature in the range of 6 K-300 K. The percentage change in $\chi^{dyn}$ is ~ 67% from ~ 60 to 6 K and is only ~ 26% from ~ 180 to 60 K and it is nearly constant from ~ 180 to 330 K. We note that $\chi^{dyn}$ remains nearly constant from room temperature to ~ 180 K, and start building up with further decrease in the temperature. Below the long-range ordering temperature (~ 60 K) it increases exponentially, clearly reflecting the signature of broken spin rotational symmetries. Spin fluctuation or magnetic excitations as a function of temperature is well characterized by the power law and it may be given as $\chi^{dyn}(T) \sim T^{\beta}$, where $\beta$ is a critical exponent. The solid red line shows the power law fit ($\beta = -0.43$). In the paramagnetic phase $\chi^{dyn}(T)$ is expected to show the temperature independent behavior owing to the spins being uncorrelated in this phase.

The magnetic specific heat ($C_m$) may be extracted from the quasi-elastic scattering part using hydrodynamic limit of the correlation function [44]. The Raman conductivity and magnetic specific heat are related as [41]:

$$\frac{\chi''(\omega)}{\omega} \propto C_m T \frac{Dk^2}{\omega^2 + (Dk^2)^2} \tag{8}$$

where $C_m$ is the magnetic specific heat and D is the thermal diffusion constant ($D = K/C_m$) with the magnetic contribution to the thermal conductivity K [41,42]. Figure 5(d) shows the spin correlation length ($\xi$) as a function of temperature in the temperature range of ~ 6-330 K. The spin correlation length $\xi(T)$ is extracted from the Lorentzian profile as described by equation (8) by taking inverse of the linewidth [41]. We observed an exponential growth of the spin correlation length below ~ 60 K, clearly reflecting the long-range ordering phase. In addition, a small change is also observed around 200 K, may be because of short-range ordering of the spins.

### 3.5 Temperature-dependent polarization of the phonon modes

One may control the polarization of the incident and scattered light and this polarized configuration may provide crucial information about the symmetry of the Raman active phonon modes. We performed the polarization dependent measurements at different temperatures by rotating the scattered light at a regular interval and fixing the incident light polarization.

Figure 6 shows the polarization dependent phonon modes intensity at five different temperatures To understand this variation of intensity as a function of polar angle, we used a semi-classical approach. According to this, intensity of the phonon modes is given as $I_{Raman} \propto |\hat{e}_s^T . R . \hat{e}_i|^2$, where 'T' indicate transpose, $\hat{e}_i$ and $\hat{e}_s$ are the unit vectors in the direction of incident and scattered light electric field. R is the Raman tensor for the corresponding phonon mode. In matrix form, the unit vectors in the direction of the incident and scattered light can be written as: $\hat{e}_i = [\cos(\theta_0) \quad \sin(\theta_0) \quad 0]; \hat{e}_s = [\cos(\theta_0 + \theta) \quad \sin(\theta_0 + \theta) \quad 0]$, where '$\theta$' is the relative angle between $\hat{e}_i$ and $\hat{e}_s$, while $\theta_0$ is the angle of incident light from the X-axis as per the schematic shown in the Fig. 4(d). The Raman tensors for the phonon modes with symmetries $A_g$, $E_{1g}$ and $E_{2g}$ are listed in the Table-III [6]. The intensity of these modes, dependent on the angle '$\theta$' and the corresponding Raman tensors are described as:

$$I_{A_g} = |a\cos(\theta)|^2 \tag{9}$$

$$I_{E_{1g}} = |[c\cos(\theta) + d\sin(\theta)]\cos(2\theta_0) + [d\cos(\theta) - c\sin(\theta)]\sin(2\theta_0)|^2 \tag{10}$$

$$I_{E_{2g}} = |[d\cos(\theta) - c\sin(\theta)]\cos(2\theta_0) - [c\cos(\theta) + d\sin(\theta)]\sin(2\theta_0)|^2 \tag{11}$$

Where, $\theta_0$ is an arbitrary angle from the X-axis as shown in the Fig. 4(d). Without any loss of generality $\theta_0$ may be taken as zero, equations (10) and (11) becomes:

$$I_{E_{1g}} = |c\cos(\theta) + d\sin(\theta)|^2, \text{ and} \tag{12}$$

$$I_{E_{2g}} = |d\cos(\theta) - c\sin(\theta)|^2 \tag{13}$$

In Figure 6, the solid red lines show the fitted curves using the equations as discussed above, see supplementary Table-S2 for the constants extracted. Modes P1, P6, and P2, P3 are fitted with equations (12) and (13) respectively. Peaks P4 and P8 are fitted with equation (9). The $A_g$ symmetry modes P4 and P8 does not show any rotation on changing temperature and the main axis shows maximum intensity at ~ 0 and 180 degrees but the loop area in the first and fourth quadrant is smaller than the loop area in the second and third quadrant and this loop area also changes with change in the temperature. The $E_{1g}$ modes P1 and P6 shows main axis rotation with changing temperature, and at 300 K main axis shows maximum intensity at ~ 0 and 180 degrees for mode P1 while 170 and 350 degree for mode P6. As we lowered the temperature to 6 K, the main axis shows the maximum intensity at ~ 170 and 350 degrees for mode P1 while for mode P2, the main axis did not change its orientation. The $E_{2g}$ modes P2 and P3 both show very small main axis rotation with change in temperature. The maxima are along the main axis at ~170 and 350 degrees. Also, the loop area of the modes P1-P4, P6, and P8 are affected by change in temperature. We note that similar rotation has also been reported recently in other 2D magnetic systems [17-18]. The microscopic understanding of these rotation calls for a detailed theoretical study to uncover all the underlying mechanism responsible for these rotations.

**Conclusion**

In conclusion, we performed a comprehensive inelastic light (Raman) scattering measurements on the single crystals of quasi-2D ferromagnetic semiconductor $Cr_2Ge_2Te_6$. Our measurements demonstrate the weak and strong symmetry breaking, attributed to the short- and long-range ordering of the spins reflected via the anomalous hardening of the phonon modes and observed anomaly in the temperature variation of the line-widths as well as intensity of modes. Our findings suggest $Cr_2Ge_2Te_6$ to be a unique platform for exploring the tunability of the interaction strength and the inelastic scattered light via symmetry control. Our results provide a critical understanding about the underlying phenomenon of this 2D magnetic compound through lattice dynamics study and further calls for exploring similar systems both experimentally as well as theoretically for a quantitative understanding particularly the polarization dependence of the interaction strength.

**Acknowledgement:** PK acknowledge support from IIT Mandi for the financial and experimental facilities.

**References:**

[1] V. Carteaux, D. Brunet, G. Ouvrard, and G. Andre, *Crystallographic, magnetic and electronic structures of a new layered ferromagnetic compound $Cr_2Ge_2Te_6$*, J. Phys. Condens. Matter **7**, 69 (1995).

[2] C. Gong et al., *Discovery of intrinsic ferromagnetism in two-dimensional van der waals crystals*, Nature **546**, 265 (2017).

[3] H. Ji et al., *A ferromagnetic insulating substrate for the epitaxial growth of topological insulators*, J. Appl. Phys. **114**, 114907 (2013).

[4] X. Li and J. Yang, *$CrXTe_3$ (X = Si, Ge) nanosheets: two dimensional intrinsic ferromagnetic semiconductors*, J. Mater. Chem. C **2**, 7071 (2014).

[5] Y. Liu and C. Petrovic, *Critical behavior of quasi-two-dimensional semiconducting ferromagnet $Cr_2Ge_2Te_6$*, Phys. Rev. B **96**, 054406 (2017).

[6] Y Tian, M. J. Gray, H. Ji, and K. S. Burch, *Magneto-elastic coupling in a potential ferromanetic 2D atomic crystal*. 2D mater. **3**, 025035 (2016).

[7] Y. Liu et al., *Short-range crystalline order-tuned conductivity in $Cr_2Si_2Te_6$ van der waals magnetic crystals*, ACS Nano **16**, 13134 (2022).

[8] V. Carteaux, G. Ouvrard, and N. Cedex, *Magnetic structure of the new layered ferromagnetic Hexatellurosilicate $Cr_2Si_2Te_6$* , j. Magn. Mater. Chem. **94**, 127 (1991).

[9] M. G. Han, J. A. Garlow, Y. Liu, H. Zhang, J. Li, D. Dimarzio, M. W. Knight, C. Petrovic, D. Jariwala, and Y. Zhu, *Topological magnetic-spin textures in two-dimensional van der waals $Cr_2Ge_2Te_6$*, Nano Lett. **19**, 7859 (2019).

[10] H. J. Koo, R. K. Kremer, and M. H. Whangbo, *High-spin orbital interactions across van der waals gaps controlling the interlayer ferromagnetism in van der waals ferromagnets*, J. Am. Chem. Soc. **144**, 16272 (2022).

[11] N. D. Mermin and H. Wagner, *Absence of ferromagnetism or antiferromagnetism in one- or two-dimensional isotropic Heisenberg models*, Phys. Rev. Lett. **17**, 1133 (1966).

[12] S. Spachmann, S. Selter, B. Büchner, S. Aswartham, and R. Klingeler, *Strong uniaxial pressure dependencies evidencing spin-lattice coupling and spin fluctuations in $Cr_2Ge_2Te_6$*, Phys. Rev. B **107**, 184421 (2022).

[13] S. Spachmann, A. Elghandour, S. Selter, B. Büchner, S. Aswartham, and R. Klingeler, *Strong effects of uniaxial pressure and short-range correlations in $Cr_2Ge_2Te_6$*, Phys. Rev. Res. **4**, L022040 (2022).

[14] Y. Sun, W. Tong, and X. Luo, *Possible magnetic correlation above the ferromagnetic phase transition temperature in $Cr_2Ge_2Te_6$*, Phys. Chem. Chem. Phys. **21**, 25220 (2019).

[15] L. Cheng et al., *Phonon-related monochromatic THz radiation and its magneto-modulation in 2D ferromagnetic $Cr_2Ge_2Te_6$*, Adv. Sci. **9**, 1 (2022).

[16] L. J. Sandilands, Y. Tian, K. W. Plumb, Y. J. Kim, and K. S. Burch, *Scattering continuum and possible fractionalized excitations in α-$RuCl_3$*, Phys. Rev. Lett. **114**, 1 (2015).

[17] B. Huang, J. Cenker, X. Zhang, E. L. Ray, T. Song, T. Taniguchi, K. Watanabe, M. A. McGuire, D. Xiao, and X. Xu, *Tuning inelastic light scattering via symmetry control in the two-dimensional magnet $CrI_3$*, Nat. Nanotechnol. **15**, 212 (2020).

[18] V. Kumar, D. Kumar, B. Singh, Y. Shemerliuk, M. Behnami, B. Büchner, S. Aswartham, and P. Kumar, *Fluctuating fractionalized spins in quasi-two-dimensional magnetic $V_{0.85}PS_3$*, Phys. Rev. B **107**, 1 (2023).

[19] D. Kumar, B. Singh, R. Kumar, and M. Kumar, *Anisotropic electron – photon – phonon coupling in layered $MoS_2$*, (2020).

[20] B. Singh, D. Kumar, V. Kumar, M. Vogl, S. Wurmehl, S. Aswartham, B. Büchner, and P. Kumar, *Fractional spin fluctuations and quantum liquid signature in $Gd_2ZnIrO_6$*, Phys. Rev. B **104**, 134402 (2021).

[21] B. Singh, M. Vogl, S. Wurmehl, S. Aswartham, B. Büchner, and P. Kumar, *Kitaev magnetism and fractionalized excitations in double perovskite $SM_2ZnIrO_6$*, Phys. Rev. Res. **2**, 013040 (2020).

[22] Y. Sun, R. C. Xiao, G. T. Lin, R. R. Zhang, L. S. Ling, Z. W. Ma, X. Luo, W. J. Lu, Y. P. Sun, and Z. G. Sheng, *Effects of Hydrostatic Pressure on Spin-Lattice Coupling in Two-Dimensional Ferromagnetic Cr2Ge2Te6*, Appl. Phys. Lett. **112**, (2018).

[23] B. H. Zhang, Y. S. Hou, Z. Wang, and R. Q. Wu, *First-principles studies of spin-phonon coupling in monolayer $Cr_2Ge_2Te_6$*, Phys. Rev. B **100**, 224427 (2019).

[24] M. Scagliotti, M. Jouanne, M. Balkanski, G. Ouvrard, and G. Benedek, *Raman scattering in antiferromagnetic $FePS_3$ and $FePSe_3$ crystals*, Phys. Rev. B **35**, 7097 (1987).

[25] I. W. Shepherd, *Tempraure dependance of phonon Raman scattering in $FeBO_3$, $InBO_3$, and $VBO_3$: evidence for a magnetic contribution to the intensities*, Phys. Rev. B. **5**, (1972).

[26] B. Singh, S. Kumar, and P. Kumar, *Broken translational and rotational symmetries in $LiMn_{1.5}Ni_{0.5}O_4$ spinel*, J. Phys. Condens. Matter **31**, 395701(2019).

[27] D. Kumar, B. Singh, P. Kumar, V. Balakrishnan, and P. Kumar, *Thermal expansion coefficient and phonon dynamics in coexisting allotropes of monolayer $WS_2$ probed by Raman scattering*, J. Phys. Condens. Matter **31**, 505403 (2019).

[28] P. G. Kelmens, *Anharmonic decay of optical phonons*, Phys. Rev. **148**, 845 (1966).

[29] R. Katoch, C. D. Sekhar, V. Adyam, J. F. Scott, R. Gupta, and A. Garg, *Spin phonon interactions and magnetodielectric effects in multiferroic $BiFeO_3$-$PbTiO_3$*, J. Phys. Condens. Matter **28**, 075901 (2016).

[30] C. J. Fennie and K. M. Rabe, *Magnetically induced phonon anisotropy in $ZnCr_2O_4$ from first principles*, Phys. Rev. Lett. **96**, 205505 (2006).

[31] D. J. Lockwood and M. G. Cottam, *The spin-phonon interaction in $FeF_2$ and $MnF_2$ studied by Raman spectroscopy*, J. Appl. Phys. **64**, 5876 (1988).

[32] E. Granado, A. García, J. A. Sanjurjo, C. Rettori, I. Torriani, F. Prado, R. D. Sánchez, A. Caneiro, and S. B. Oseroff, *Magnetic ordering effects in the Raman spectra of $La_{1-x}Mn_{1-x}O_3$*, Phys. Rev. B. **60**, 11879 (1999).

[33] U. Fano, *Effects of Configuration Interaction on Intensities and Phase Shifts*, Phys. Rev. **124**, 1866 (1961).

[34] M. F. Limonov, M. V. Rybin, A. N. Poddubny, and Y. S. Kivshar, *Fano resonances in photonics*, Nat. Photonics **11**, 543 (2017).

[35] A. E. Miroshnichenko, S. Flach, and Y. S. Kivshar, *Fano resonances in nanoscale structures*, Rev. Mod. Phys. **82**, 2257 (2010).

[36] T. Moriya, *Thory of light scattering by magnetic crystals,* Journal of the physical society of Japan, **23** (1967).

[37] R. P. Wang, G. Xu, and P. Jin, *Size dependence of electron-phonon coupling in ZnO nanowires*, Phys. Rev. B . **69**, 5 (2004).

[38] W. Zhang, T. J. A. Craddock, Y. Li, M. Swartzlander, R. R. Alfano, and L. Shi, *Fano resonance line shapes in the Raman spectra of tubulin and microtubules reveal quantum effects*, Biophys. Reports **2**, (2022).

[39] W. Zhang, T. J. A. Craddock, Y. Li, M. Swartzlander, R. R. Alfano, and L. Shi, *Fano Resonance Line Shapes in the Raman Spectra of Tubulin and Microtubules Reveal Quantum Effects*, Biophys. Reports **2**, (2022).

[40] C. I. Medel-Ruiz, H. P. Ladrón de Guevara, J. R. Molina-Contreras, and C. Frausto-Reyes, *Fano effect in resonant Raman spectrum of CdTe*, Solid State Commun. **312**, 1 (2020).

[41] Y. Choi, S. Lee, J. H. Lee, S. Lee, M. J. Seong, and K. Y. Choi, *Bosonic spinons in anisotropic triangular antiferromagnets*, Nat. Commun. **12**, 1 (2021).

[42] A. Glamazda, P. Lemmens, S.H. Do, Y.S. Choi, and K.Y. Choi, *Raman spectroscopic signature of fractionalized excitations in the harmonic-honeycomb iridates β- and γ-$Li_2IrO_3$,* Nat. commun. **7**, 12286 (2016).

[43] S. Yoon, K. Y. Choi, and H. Nam, *Raman Spectroscopic Probe of the Magnetic Specific Heat in Quantum Magnets*, J. Korean Phys. Soc. **77**, 138 (2020).

[44] B. I. Halperin and P. C. Hohenberg, *Hydrodynamic Theory of Spinwaves*, J. Appl. Phys. **40**, 1554 (1969).

**Figure Captions:**

**Figure 1: (a)** Raman spectrum of the single crystal of $Cr_2Ge_2Te_6$ in the spectral range of 65 to 500 cm$^{-1}$ recorded at 6 K. Insets in the green-shaded area are the amplified spectra in the spectral range of 65 to 120 cm$^{-1}$ (left side) and 200 to 500 cm$^{-1}$ (right side). **(b)** Shows the temperature evolution of the Raman spectrum in the spectral range of 5 to 500 cm$^{-1}$. Left inset show temperature evolution of the most prominent peaks i.e. P3 and P4. Right inset shows the spectra up to ~ 2500 cm$^{-1}$. **(c)** Shows the crystal structure of $Cr_2Ge_2Te_6$, plotted using VESTA, as viewed along the $c$-axis (top-view). The unit cell is shown by the black frame, and **(d)** shows the structure along the arbitrary direction.

**Figure 2: (a)** and **(b)** Show temperature dependence of the frequency and FWHM of the phonon mods P1-P3, respectively; **(c)** and **(d)** Show temperature dependence of the frequency and FWHM of the phonon modes P4, P6, and P8, respectively, for $Cr_2Ge_2Te_6$. (The solid red line shows a three-phonon fitting in the temperature range 60 to 330K and the green lines in the temperature range 6 to 60K are a guide to eye drawn by extrapolation. In the inset, the solid blue line shows the spin-phonon couling fitting for the temperature range of 6-60 K ). $T_c$ (~ 60 K) and $T^*$ (~ 180 K) represents the long and short range ordering temperature, respectively. **(e)** Shows the intensity variation as a function of temperature (6-330 K) for phonon modes P1-P4, and P8 respectively. Red lines are guide to the eye.

**Figure 3: (a)** Shows the temperature dependence of the linear thermal expansion coefficient for the phonon modes P1-P3, and **(b)** shows the temperature dependence of the linear thermal expansion coefficient for the phonon modes P4, P6, and P8; respectively in the temperature range 60-330 K with three values of Gruneisen parameters $\gamma$ = 1, 2, 3.

**Figure 4: (a)** Evolution of a cumulative fit of the Fano peak (P3) for different $q$ parameters and temperatures. **(b)** Temperature dependence of the asymmetry parameter $1/|q|$ in the temperature range 6 to 330 K. The solid semi-transparent blue line is drawn as guide to the eye below 60 K. **(c)** Shows the polarization dependence of the asymmetry parameter, $1/|q|$, at temperature 6K, 45K, 150K, 250K, and 300K. Polarization dependent measuremenst are performed by fixing the direction of polarization of the incident light and rotating the scattered light polarization **(d)** Shows schematic for the unit vectors ($\hat{e}_i$ and $\hat{e}_s$) for the incident and scattered light. $\hat{K}_i$ and $\hat{K}_s$ shows the direction of propagation of the incident and scattered light.

**Figure 5: (a)** Temperature evolution of the Raman response $\chi''(\omega,T)$) [measured raw Raman intensity/1 + n(ω)]). **(b)** Temperature dependence of the Raman conductivity $\chi''(\omega,T)/\omega$. **(c)** Temperature dependence of the dynamic Raman susceptibility $\chi$. The solid red line shows the fitting for power law. **(d)** Shows the temperature dependence of the spin correlation length ($\xi$) extracted from the Raman conductivity. The inset shows the amplified region in the temperature range of ~ 140-330 K. The semi-transparent red line is drawn as guide to the eye.

**Figure 6:** Polarization dependent intensity of phonon modes P1-P4, P6, and P8 at different temperatures 6K, 45K, 150K, 250K, and 300K, respectively with scattering light rotated configuration.

**Figure 1:**

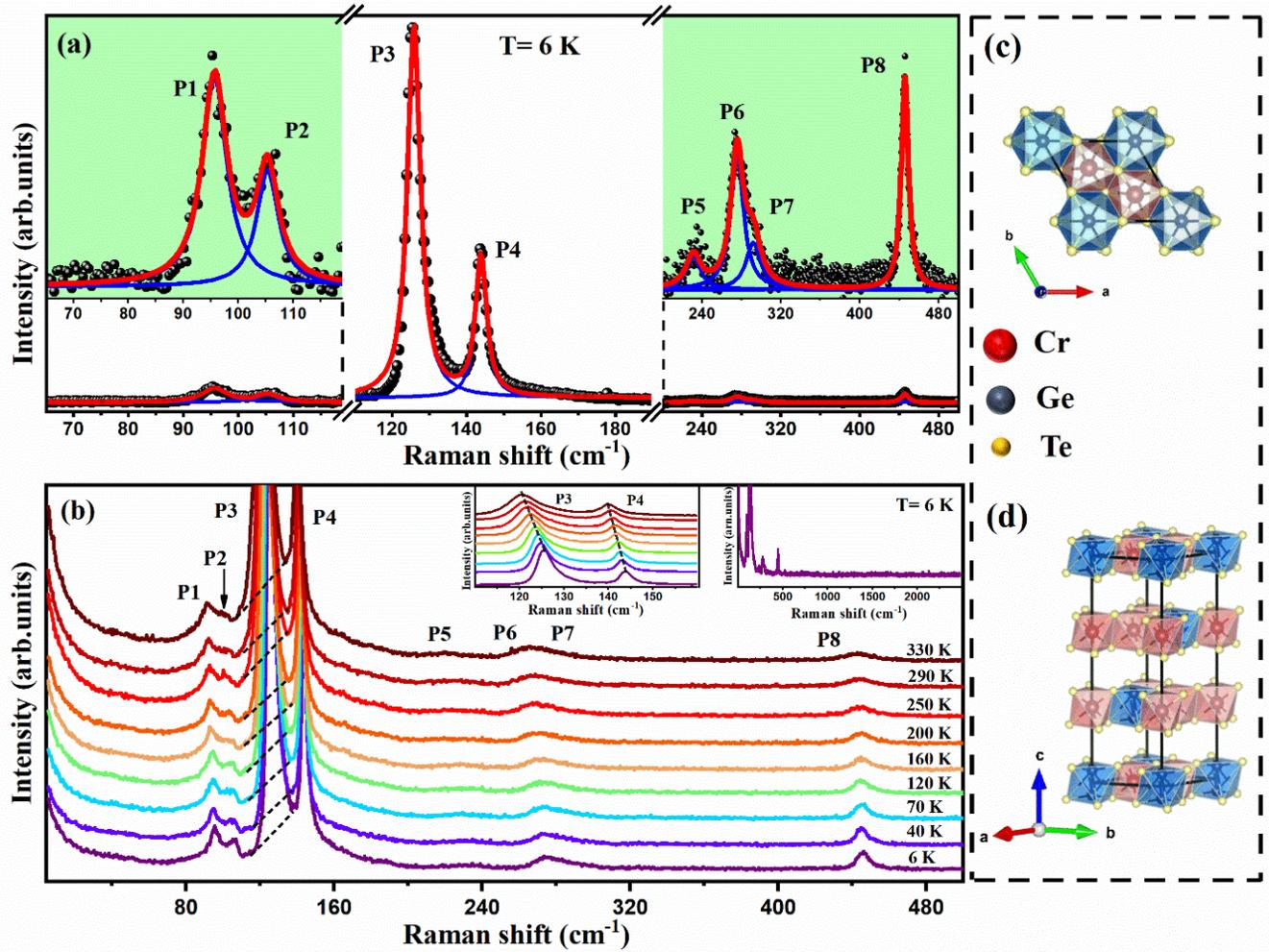

**Figure 2:**

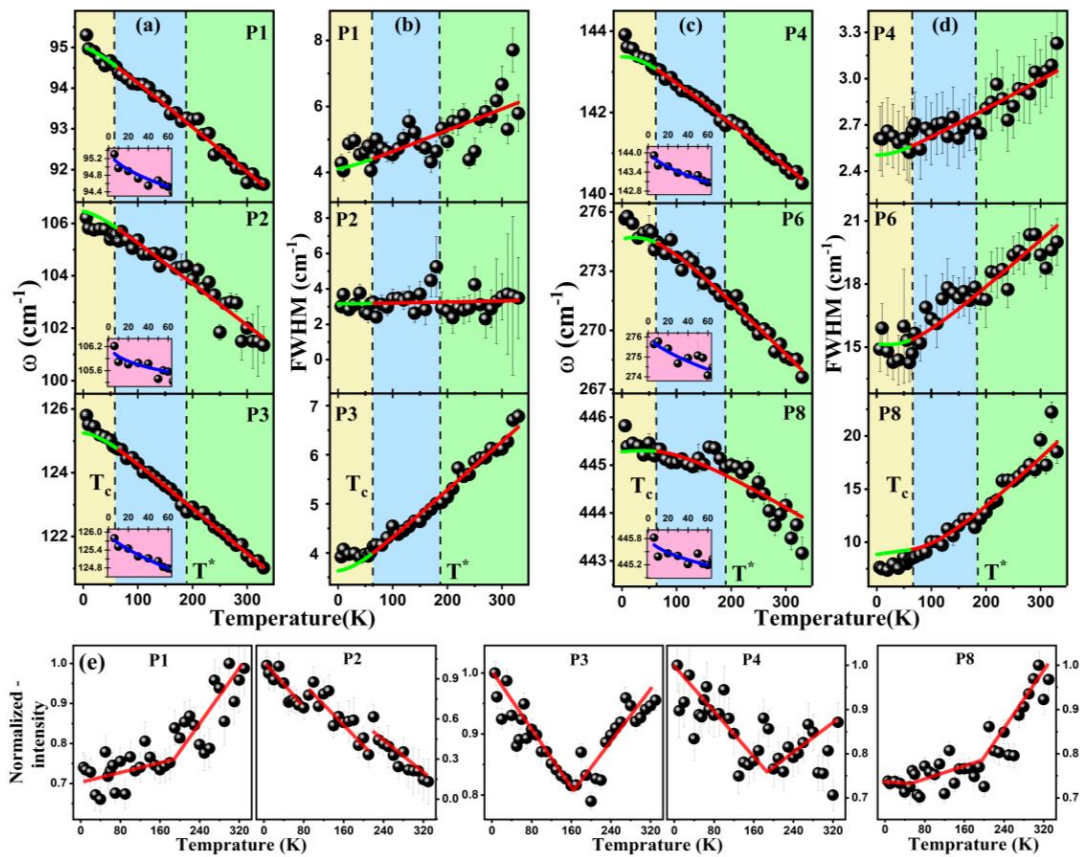
24

**Figure 3:**

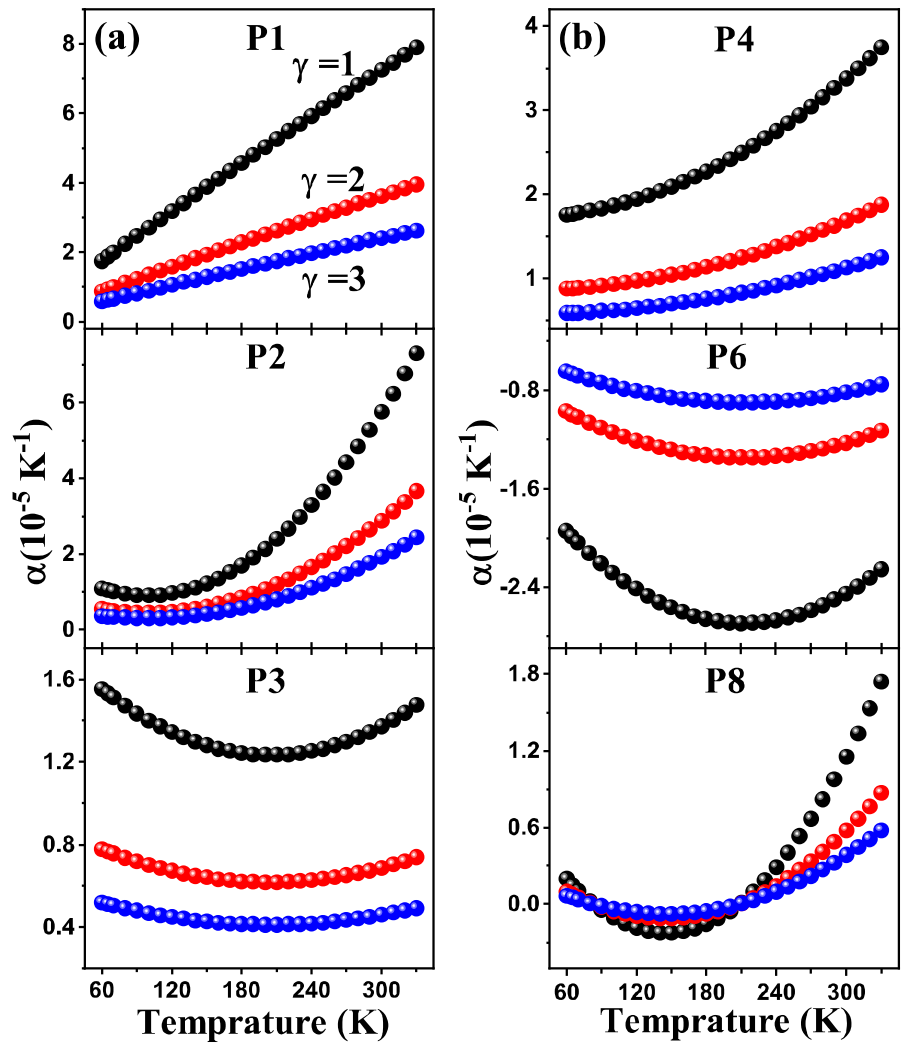

**Figure 4:**

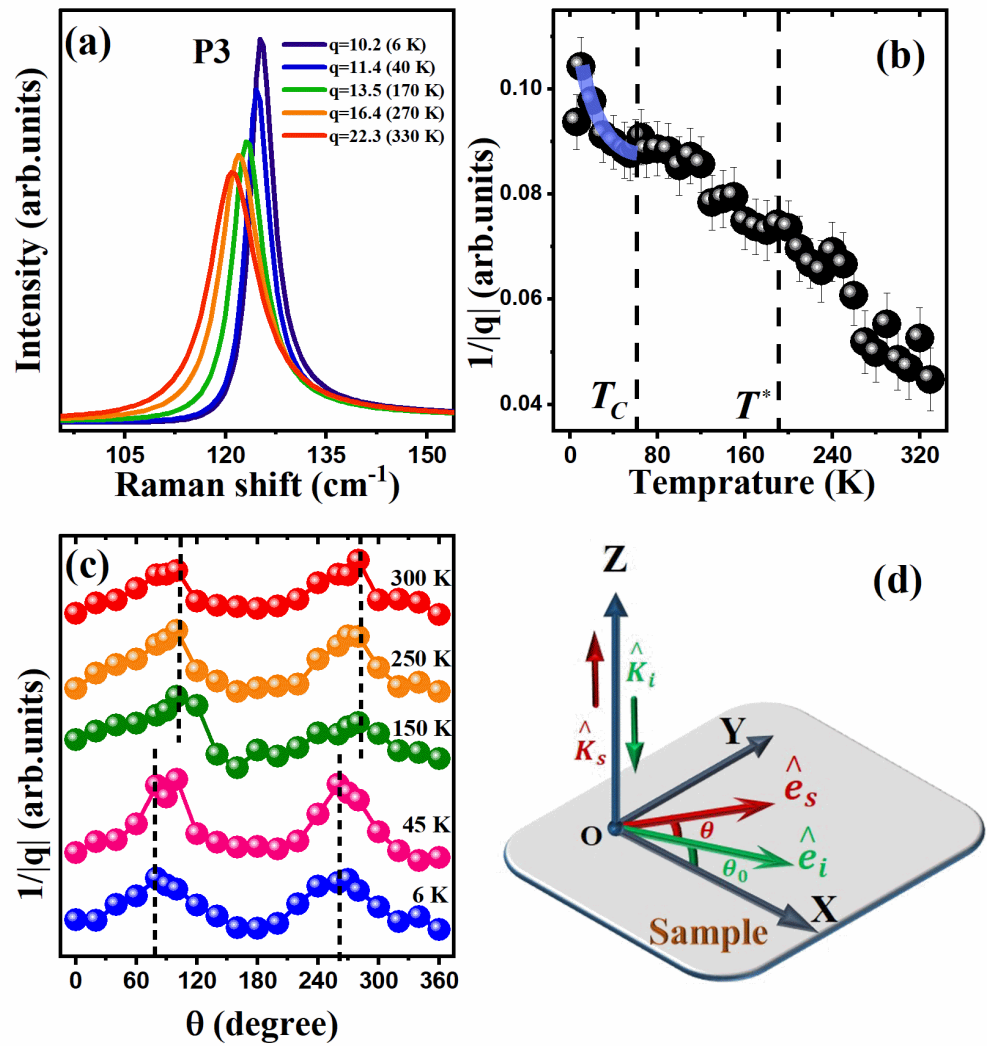

**Figure 5:**

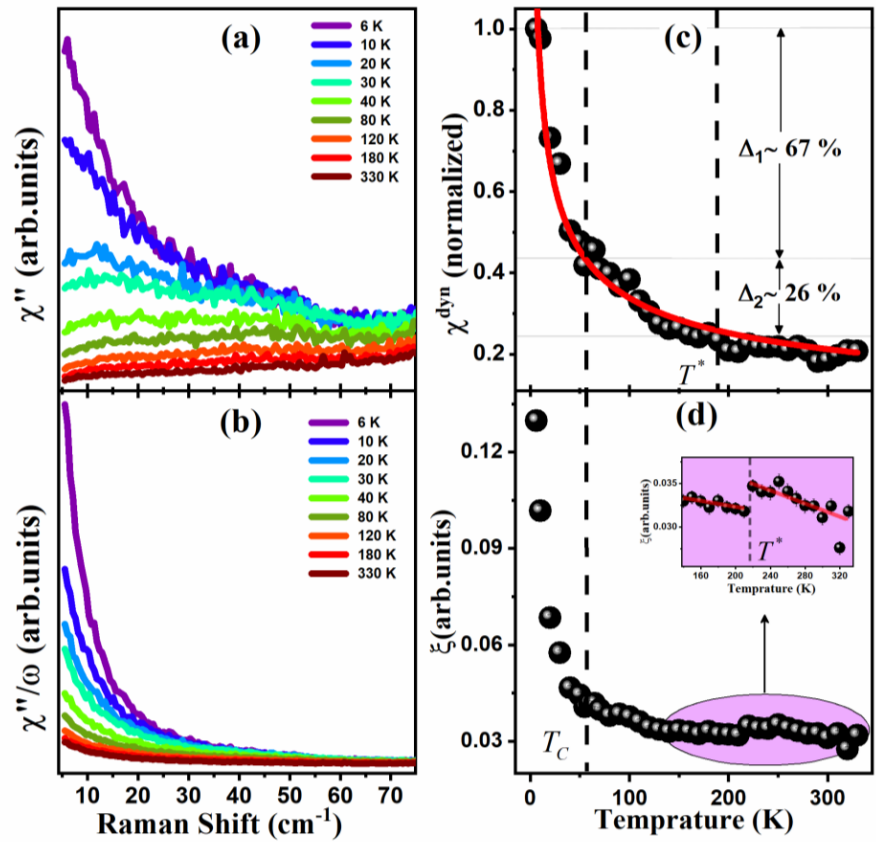

**Figure 6:**

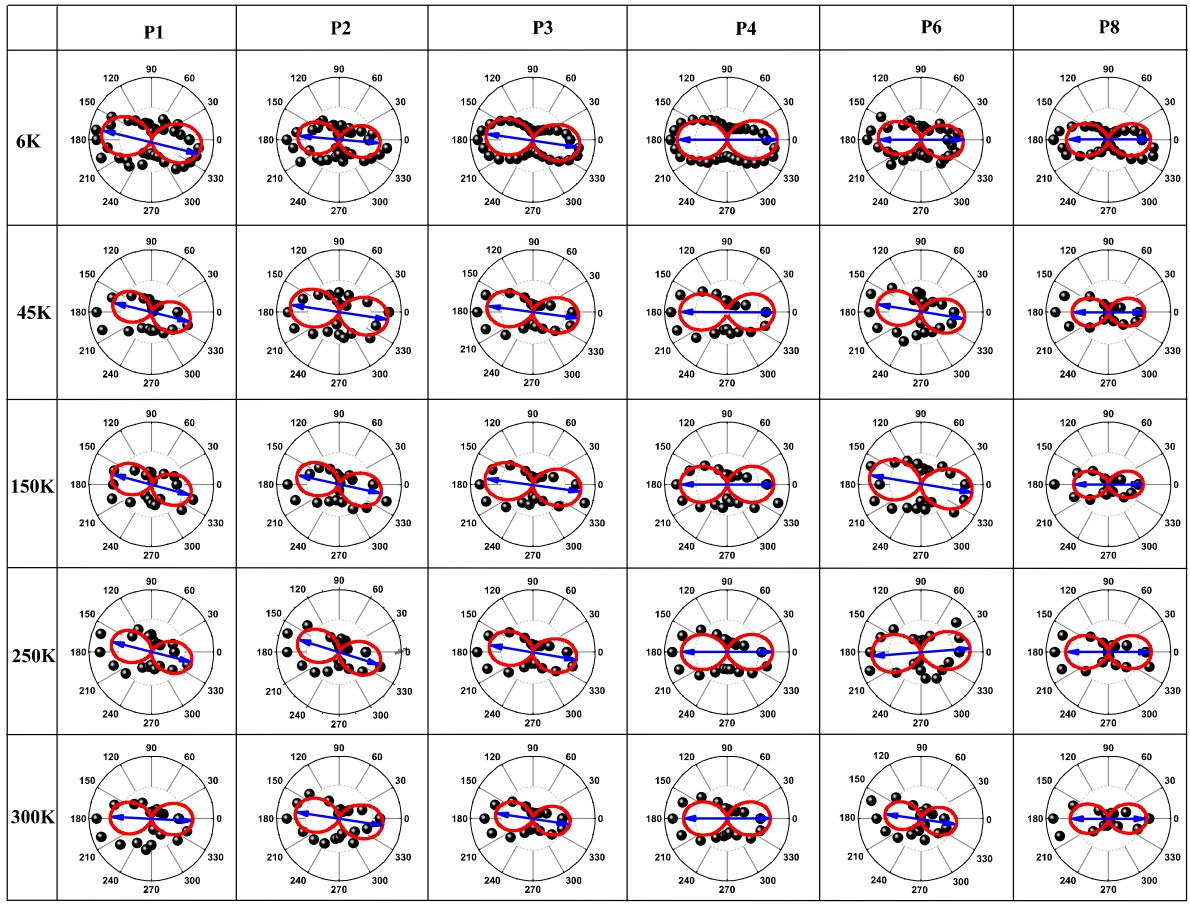

**Table-I.** List of the fitting parameters corresponding to the phonon modes in $Cr_2Ge_2Te_6$, fitted using the three-phonon fitting model in the temperature range of ~ 60 to 330 K. The units are in $cm^{-1}$.

| Mode assignment | $\omega_0$ | A | $\Gamma_0$ | C |
|---|---|---|---|---|
| **P1** ($A_g$) | 95.3 ± 0.06 | - 0.4 ± 0.01 | 4.0 ± 0.30 | 0.2 ± 0.04 |
| **P2** ($E_{2g}$) | 106.9 ± 0.20 | - 0.6 ± 0.04 | 3.1 ± 0.31 | 0.02 ± 0.06 |
| **P3** ($E_{2g}$) | 125.8 ± 0.04 | - 0.7 ± 0.01 | 3.3 ± 0.06 | 0.5 ± 0.01 |
| **P4** ($A_g$) | 143.9 ± 0.04 | - 0.5 ± 0.01 | 2.4 ± 0.04 | 1.0 ± 0.01 |
| **P6** ($E_{1g}$) | 277.5 ± 0.18 | - 2.8 ± 0.08 | 12.8 ± 0.32 | 2.3 ± 0.18 |
| **P8** ($A_g$) | 446.5 ± 0.17 | - 1.1 ± 0.13 | 0.8 ± 0.00 | 8.4 ± 0.11 |

**Table-II:** Spin-phonon coupling fitting parameters extracted using expression described in the text.

| Mode assignment | $\lambda$ | $\eta$ |
|---|---|---|
| **P1** | -0.40 ± 0.2 | 0.55 ± 0.47 |
| **P2** | -0.33 ± 0.2 | 1.0 ± 0.8 |
| **P3** | -0.52 ± 0.1 | 0.71 ± 0.2 |
| **P4** | -0.48 ± 0.2 | 0.54 ± 0.3 |
| **P6** | -0.75 ± 0.5 | 0.68 ± 0.9 |
| **P8** | -0.31 ± 0.2 | 0.5 ± 0.7 |

**Table-III:** Wyckoff positions and irreducible representations of phonon modes for $Cr_2Ge_2Te_6$ in the hexagonal ($C_{3i}$; space group No.148) phase.

| Atom | Wyckoff position | Γ-point mode decomposition | Raman tensor |
|---|---|---|---|
| Te | 18f | $3A_g + 3A_u + 3E_g + 3E_u$ | $A_g = \begin{pmatrix} a & 0 & 0 \\ 0 & a & 0 \\ 0 & 0 & b \end{pmatrix}$ |
| Cr | 6c | $A_g + A_u + E_g + E_u$ | $E_{1g} = \begin{pmatrix} c & d & e \\ d & -c & f \\ e & f & 0 \end{pmatrix}$ |
| Ge | 6c | $A_g + A_u + E_g + E_u$ | $E_{2g} = \begin{pmatrix} d & -c & -f \\ -c & -d & e \\ -f & e & 0 \end{pmatrix}$ |
| Raman and infrared active modes | | $\Gamma_{IR} = 4A_u + 4E_u$ $\Gamma_{Raman} = 5A_g + 5E_g$ | |

# Supplementary information:

## Broken weak and strong spin rotational symmetries and tunable interaction between phonon and the continuum in $Cr_2Ge_2Te_6$

Atul G. Chakkar[*], Deepu Kumar, Pradeep Kumar[#]

*School of Physical Sciences, Indian Institute of Technology Mandi, 175005, India*

[*]E-mail: atulchakkar16@gmail.com
[#]E-mail: pkumar@iitmandi.ac.in
**Table-S1:** The extracted constant parameters by fitting with three phonon + TEC model as described in the text.

|     | $b_0$ ($\times 10^{-5}$) | $b_1$ ($\times 10^{-7}$) | $b_2$ ($\times 10^{-10}$) |
|-----|--------------------------|--------------------------|---------------------------|
| **P1** | 0.3034 | 2.4572  | -4.7956 |
| **P2** | 2.0557 | -2.3424 | 11.9175 |
| **P3** | 2.2086 | -0.5648 | 1.4133  |
| **P4** | 1.7237 | -0.6002 | 2.0381  |
| **P6** | 1.9205 | 0.8011  | -1.3460 |
| **P8** | 0.9987 | -1.6753 | 5.7713  |

**Table-S2:** The extracted constant parameters for the polar plot at five temperatures using equations as mentioned in the text.

| Peak | P1 | | P2 | | P3 | | P4 | P6 | | P8 |
|---|---|---|---|---|---|---|---|---|---|---|
| constant | c | d | c | d | c | d | a | c | d | a |
| **6K** | 36.1 | -4.7 | 2.1 | 24.3 | 17.4 | 149.1 | 83.6 | 45.1 | -4.0 | 43.0 |
| **45K** | 33.1 | -8.3 | 3.7 | 23.8 | 19.9 | 153.3 | 85.7 | 48.4 | -8.1 | 44.4 |
| **150K** | 32.1 | -8.2 | 5.6 | 23.9 | 20.3 | 160.6 | 88.4 | 49.1 | -8.3 | 42.2 |
| **250K** | 29.4 | -7.4 | 5.5 | 19.5 | 25.4 | 160.4 | 87.4 | 52.4 | -2.8 | 46.1 |
| **300K** | 28.1 | -0.8 | 3.0 | 21.2 | 19.6 | 174.9 | 93.1 | 37.5 | -5.6 | 38.2 |